# DiffGame: Game-based mathematics learning for physics

Mads Kock Pedersen[a], Anette Svenningsen[a], Niels Bonderup Dohn[b], Andreas Lieberoth[a,b], and Jacob Sherson[a]1

[a]Center for Community Driven Research, Aarhus University, Ny Munkegade 120, 8000 Aarhus C, Denmark
[b]Danish School of Education, Aarhus University, Niels Juels Gade 84, 8200 Aarhus N, Denmark

**Abstract**

Differentiation is a mathematical skill applied throughout science in order to describe the change of a function with respect to a dependent variable. Thus, an intuitive understanding of differentiation is necessary to work with the mathematical frameworks used to describe physical systems in the higher levels of education. In order to obtain this intuition repeated practice is required. This paper presents the development of DiffGame, which consists of a series of exercises that introduce the basic principles of differentiation for high-school students through game-like elements. DiffGame have been tested with 117 first-year students from a single Danish high school, who did not have any prior training in differentiation. The students' learning was assessed by the data obtained directly from DiffGame. The test demonstrated the efficacy of DiffGame, since students at all levels demonstrate a learning gain. In contrast to previous studies demonstrating most learning in the lower tier of students, the middle tier of students (based on overall performance) exhibits the largest learning gains.

**Introduction**

Differentiation is a mathematical skill that determines the change in a function with respect to a change in a dependent variable. Differentiation has a prominent role in the mathematics that describes physical systems. As an example, the resulting force acting on an object at any given time can be determined by differentiating its velocity with respect to time and thereby obtain the acceleration, which is proportional to the force according to Newton's Second law. Thus, knowing how to differentiate is crucial for any student who wants to proceed with the higher levels of education in science. In order to obtain an intuition about the basic principles of differentiation, repeated practice in solving differentiation-problems is necessary. Students are, however, usually first introduced to the concepts through lectures where the math is presented in an abstract format. Earlier studies have shown an improved learning process by letting students first investigate and try to solve problems within a subject on their own before exposing them to the formal knowledge that enables novices to form an initial understanding of the subject (Kapur et al., 2010).

---

* Corresponding author.
  *E-mail address:* sherson@phys.au.dk

One way to facilitate students' independent investigation of differentiation-problems is through games. In an educational context games have the ability to make learning more entertaining and motivate students without compromising the learning (Muntean, 2011). Abstract content which could otherwise be incomprehensible can in games be conveyed though simple visual representations and interactions (Bjælde et al., 2014; Magnusen et al., 2014; Pedersen et al., 2016). The simple visual representations combined with the game elements allow students to obtain an intuition even for systems which do not have any analogies to their daily life (Lieberoth et al., 2015).

Games can be designed such that students are presented with problems in an environment that allows them to make mistakes and reflect upon them. The flow-in-game infused learning can be designed to adapt to the individual students' level of skill in order to provide training in the most beneficial areas. Such dynamically differentiated learning (Algozzine & Anderson, 2007) is enabled by giving students, who have troubles with a specific type of problems, extra opportunities to train, while allowing those, who have mastered the problems, to proceed. Thus, students will mostly work in their zone of proximal development (Vygotsky, 1978); continually challenged to the border of their current abilities but with scaffolds like tutoring and feedback in place to keep the learning trajectory moving forward (Wood, Bruner, & Ross, 1976). These elements align well with the visible learning theory (Hattie & Yates, 2014); the games welcome errors as learning opportunities, the students current abilities are assessed and determine the next step in the learning process, the students have a clear goal to fulfill, and the games can be built such that both teachers and students can easily follow the progress.

This paper presents the development of DiffGame, which consists of a series of exercises that introduce the basic principles of differentiation for high-school students through game-like elements. The games were developed and tested in a blended-learning environment, in which the games were deployed in a classroom with a teacher and instructors at hand to help and guide the sessions. The aim was to develop the games such that they required no previous knowledge about differentiation, but it could also be used to refresh students' differentiation knowledge at later stages of their education.

Previous studies (Bjælde et al., 2014; Adams et al., 2008) have showed that it is often the lower tier of students, who learn the most from games. However, it remains an open question whether these results are just a case of diminishing returns (the more you know the less there is to learn), or if gamified teaching appeals to a different segment of students than traditional lectures. Thus, the students are divided into lower, middle, or upper tiers of students based on overall performance in DiffGame and their learning gains are compared.

Another point of interest is how the DiffGame motivated students, and to what degree such motivation predicted performance. Motivation refers to the process whereby goal-directed activities are instigated and sustained by intrinsic or extrinsic dynamics (Lepper et al., 1973; Lieberoth et al., 2014; Ryan & Deci, 2000). Intrinsic motivation refers to motivation to engage in an activity for its own sake. When students are intrinsically motivated, they work on tasks because they enjoy the activity and find it meaningful. In contrast, extrinsic motivation is based on other reasons, such as outside demands or economic and symbolic rewards (Ryan & Deci, 2000). In most cases learning trajectories are supported by a little bit of both. Since intrinsic motivation is thought to have a positive impact on students (Deci et al., 1991), the potential relationship between high levels of intrinsic motivation and higher levels of performance in DiffGame is examined.

**DiffGame**

Differentiation is an important part of the Danish high school curriculum. Differentiation is always introduced through visual interpretation as the slope of a curve at a particular point. However, most teaching on the subject then proceeds quickly to the more formalistic rules for deriving various analytic functions. DiffGame delves much longer at the visual representation and use it as a tool for the students to explore the formalistic rules. The ability to categorize a function from its visual or analytical form is a crucial ability in order to identify the patterns of the derivative of functions. The rules that govern how to differentiate a function depend on the function category. DiffGame included six distinct categories of functions across all the games in increasing order of difficulty; linear functions, 2nd and 3rd order polynomials, exponentials, power functions, logarithmic functions, and trigonometric functions. A seventh category called Advanced included sums of functions from the other six categories and were only included as an extracurricular mode. The first version of DiffGame included six interactive exercises, which were hosted on a website with links to the previous page and next page. However, based on the pilot tests three of the exercises were removed in order to streamline the overall flow.

In the introductory Recognize Function the students are challenged with the task of identifying a function. The game has two operational modes; either the graph of the function is displayed and the student then has to pick the corresponding analytical expression given four choices as seen in Figure 1.a, or the analytical expression for the function is given and the student has to pick the correct graph given four choices. When submitting their answers the students get immediate feedback presenting whether their answer was correct or not, and what the correct answer is. In the initial training part of the activity the student can pick which of the seven categories to include in their training. This freedom of choice is also meant as an easy tool for the teachers to deliver level-differentiated teaching. When the students have answered three correct in row, indicated by three bars that change to green when a streak of correct answers is obtained, the student will be sent to the next category. If the student does not get three correct in a row within ten tries, he/she is sent to the next category. Below the activity-frame is a small description of how to identify the specific type of function from the graph. After all chosen categories are completed the student is given ten randomly drawn questions.

In Sign of slope the students are challenged to identify which of three intervals fulfils a specific condition for the slope (positive, negative, or zero). A random graph is shown and three intervals are marked as seen in Figure 1.b. When students select an interval, they are given feedback on whether they answered correctly or incorrectly.

Finally, in the Differentiation Game the students are shown the analytical expression and the graph of a function, and need to identify the analytical expression for the derivative given four choices. When students submit their answers, they get immediate feedback on whether the question was correct or incorrect. When students click on a point on the graph of the original function, seen to the left on Figure 1.c, the tangent in that point is plotted and the slope of the function in that point - i.e. the derivative in that point – is plotted into a new coordinate system, as seen on the right on Figure 1.c. This access to dynamic "hints" is expected to help the students learn how to find the derivative even in cases where the analytic form of the derivative of the particular function has not yet been taught. Similar to Recognize Functions the game consists of two parts. In the first training part the students are given basic training, before they proceed onto the game mode, in which a total of 20 exercises must be completed with less than three incorrect answers, and with less than a total of 50 hints, i.e. clicks on the original function to get help. If the student gets three wrong or answers all 20 exercises, the game mode stops and a score is generated based on the number of correct answers, time used on each task, and the amount of unused hints still left.

**Methods**

DiffGame has been through two tests with a total of 168 students. In the first pilot DiffGame was given to a total of 51 Grade 11 students (36 males, age 16-19) from two different Danish high schools who were just about to commence classes in differentiation. The pilot test was conducted as two hour sessions, in which the students were supposed to work through DiffGame on their own. After this, all students answered a multiple-choice test on a 1-6 scale from strongly disagree to strongly agree with the 6 items from the value/usefulness and 7 items from the interest/enjoyment scales of the Intrinsic Motivation Inventory. A focus-group interview with selected students was also conducted.

Based on the observations made during the pilot, three of the frames were removed from the track and the accompanying texts were shortened and sharpened. Most importantly, a critical software bug was identified and fixed. In the second test, 117 Grade 10 students (87 of 88 are males, age 16-23) from a single high school participated, again with no prior training in differentiation. However, since they had no prior experiences with differentiation the categories were limited to linear functions and polynomials. This second test was also a two hour session, in which they went through the track and answered the survey.

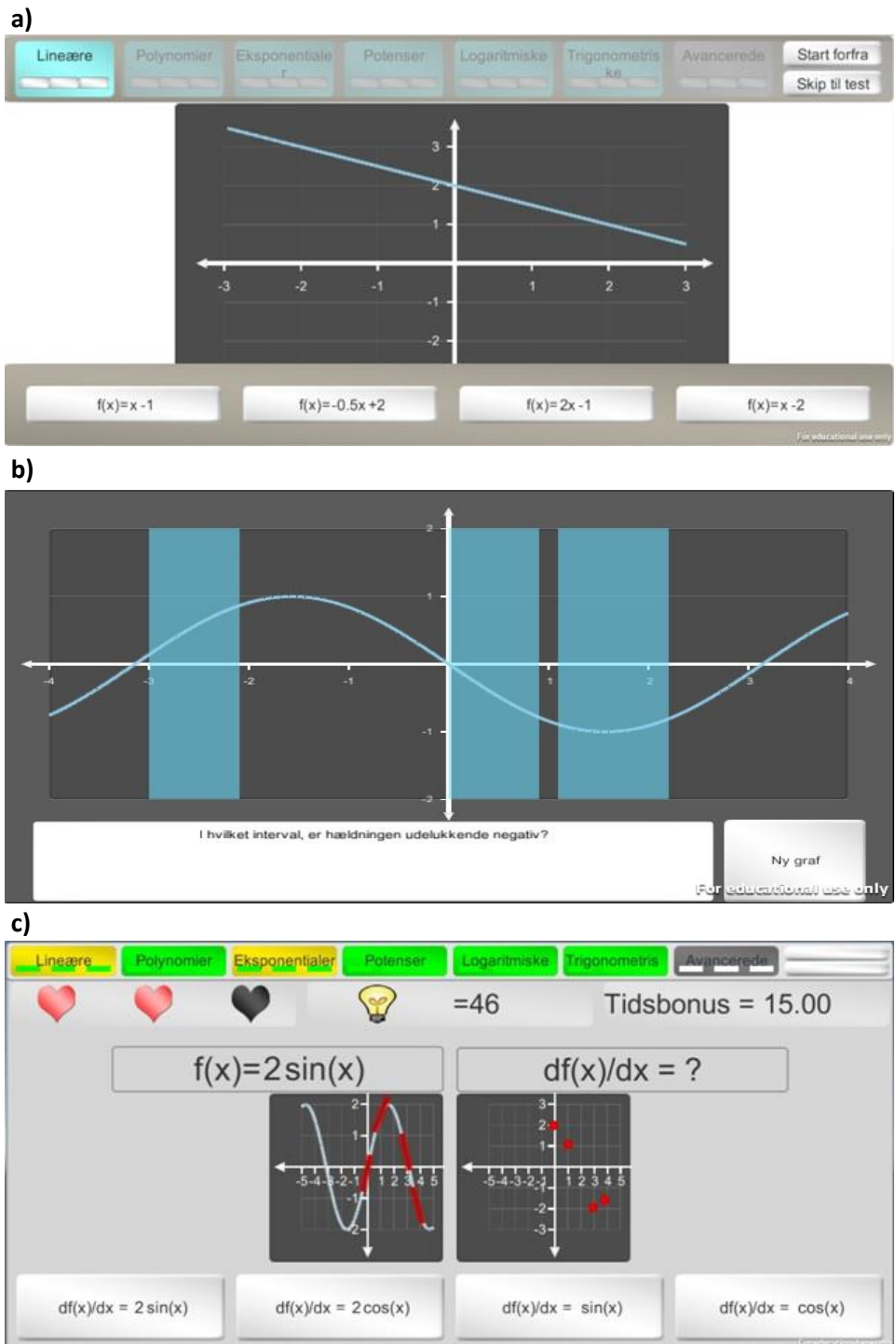

Figure 1. Screenshots from the various frames in DiffGame. a) Recognize function with a given graph and unknown analytical expression. b) Sign of slope with the challenge of identifying in which of the intervals is the slope only negative. c) Differentiation Game where students select the correct derivative of the function to the left. By clicking the graph, students get the slope of the function in that point, and from this identify the derivative.

During the pilot test it became apparent that the differentiation game contained a critical software bug; the number 32 appeared in random places and obscured the data. Thus, the analysis focuses solely on the results from the second and larger test below.

In order to compare the students' learning and relative skill, the assessment of the students was derived from game data which minimizes the divide between testing the students and their active learning paths. All responses were time stamped and stored in a database allowing the examination of all students' progression through DiffGame. This enables the collective assessment of both DiffGame's learning curve and the students' abilities as learning trajectories flowed forward. The difference between each student's first three and last three answers in each category of exercises, plus each student's percentage of correct answers within each category, were extracted in order to assess learning and relative skill respectively.

**Results**

The students' learning is quantified by each student's gain (Bjælde et al., 2014) i.e. the fraction of the potential learning that was realized through the game

$$Gain = \frac{N_{\text{correct last}} - N_{\text{correct first}}}{N_{\text{exercies}} - N_{\text{correct first}}},$$

where $N_{\text{correct last}}$ and $N_{\text{correct first}}$ are the number of correct answers in the last three and the first three exercises, and $N_{\text{exercises}}$ is the number of compared exercises. The average gain for all 117 students was large (0.65 with a 0.12 standard-error-of-mean. Figure 2.a compares the difference in correct answers between the first three and last three exercises between 20% percentiles of students based on their overall percentage of correct answers.

The average gains within the percentiles are displayed above the average number of correct answers in Figure 2.a, and all percentiles demonstrates a large gain. Figure 2.b. access learning by comparing the percentage of correct answers in the training and the game.

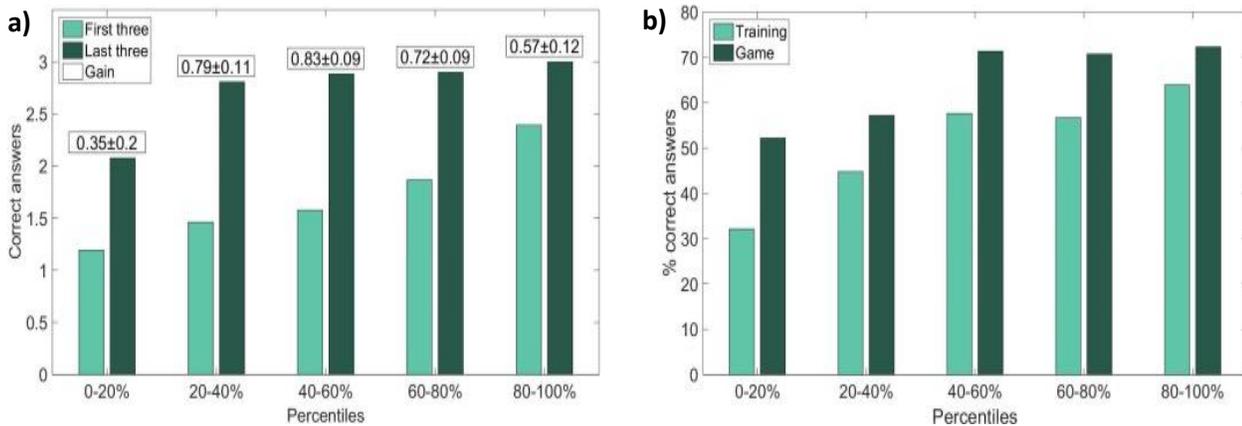

Figure 2. Students' improvement in the differentiation game. a) The average number of correct answers in the first three and last three exercises in the training for each percentile of students. The average gains for the percentiles and their standard-error-of-mean are displayed above the bars. b) The average percentage of correct answers in the training and the game for each percentile of students.

60 of the 117 students completed the Intrinsic Motivation Inventory surveys. The enjoyment (M = 4.00, SD = 0.93) and value (M = 4.27, SD = 0.93) scales yielded good Cronbach alphas of 0.91 and 0.90 respectively. In order to explore the relationship between intrinsic motivation in the situation on one hand, and levels of performance on the other, Spearman correlations between Intrinsic Motivation Inventory scores with the percentage of correct answers are computed, which for both the enjoyment and value scales were insignificant.

## Discussions

The second test demonstrated a gain from both the weakest (0.35) and the strongest (0.57) students with the middle tier displaying the largest gain (0.83), and the same trend manifested in the training to game data, this contradicts some previous studies (Bjælde et al. 2014, Adams et al. 2008), where the weakest students demonstrate the largest gains. However, it is an open question whether the large gains stems purely from a training effect familiarizing students with the format rather than creating transferable skills and understandings. The training effect might favor students with some modicum of intuitions about the math involved while stifling those who had to deal with both a new learning format and a task which would have been difficult for them no matter what. In conversations with teachers, they said that the curriculum of DiffGame was equivalent to 4-6 weeks of traditional classes. Thus, the predicted outcome of a controlled test is that the DiffGame will yield large immediate gains, but also that this will take some work to transfer and align with the other modes of learning.

The intrinsic motivation scales showed that the students generally found the games to be fun and useful, but since the tests could not establish a relationship between level of performance and high levels of intrinsic motivation, it appears that extrinsic motivation was also a driving force behind participation (similarly to conclusions drawn by Mekler et al., 2015) even if game-inspired design elements like interactivity, feedback and clear goals scaffold students to work on every tasks.

## Conclusion

The main result from the pilot test was the realization that three of the original six frames in DiffGame did not fulfill their purpose due to a common pattern: The lack of a well-defined objective was common for the three frames, and during the pilot the students got stuck in these, because they did not know when to proceed. Thus, in order to streamline DiffGame, these frames were removed until they are redesigned with clearer objectives.

The second test demonstrated the viability of DiffGame. Learning gains were observed for all students, with the middle tier of students demonstrating the largest gain. The students in general found the DiffGame to be both enjoyable and valuable, even though no connections could be established between these intrinsic motivations and performance.

A future study with a control group receiving a traditional lecture and training exercises would be essential to determine the comparable efficiency of the DiffGame in terms of both motivation and learning curve. However, given the small amount of time needed for the activity and the broad scope of the material covered DiffGame is still a versatile tool for high school and university education enabling both fun introduction to and easy repetition of the topic of differentiation as well as a flexible teacher tool for skill-level differentiation.

## Acknowledgements

The authors would like to thank the John Templeton Foundation and the Lundbeck Foundation for financial support.

## References

Adams, W. K., Reid, S., LeMaster, R., McKagan, S. B., Perkins, K. K., Dubson, M., Wieman, C. E. (2008). A study of educational simulations Part I - Engagement and learning. *Journal of Interactive Learning Research*, *19*, 397.

Algozzine, B. & Anderson K. M. (2007). Tips for Teaching: Differentiating Instruction to Include All Students. *Preventing School Failure, 51*, 49-54.

Bjælde, O. E., Pedersen, M. K., & Sherson, J. (2014). Gamification of Quantum Mechanics Teaching. *World Conference on E-Learning in Corporate, Government, Healthcare, and Higher Education*, 218-222.

Deci, E. L., Vallerand, R. J., Pelletier, L. G., & Ryan, R. M. (1991). Motivation and education: the self-determination perspective. *Educational Psychologist, 26*, 325–346.

Hattie, J. & Yates, G. (2014). *Visible Learning and the Science of How We Learn*. New York: Routledge.

Kapur, M., Dickson, L., & Yhing, T. P. (2010). Productive Failure in Mathematical Problem Solving. *Instructional Science, 38*, 523-550.

Lepper, M. R., Greene, D., & Nisbett, R. E. (1973). Undermining Children's Intrinsic Interest with Extrinsic Reward: a test of the "overjustification" hypothesis. *Journal of Personality and Social Psychology, 28*, 129

Lieberoth, A., Pedersen, M. K., Marin, A., Planke, T., & Sherson, J. F. (2014). Getting Humans to Do Quantum Optimization - User Acquisition, Engagement and Early Results from the Citizen Cyberscience project Quantum Moves. *Human Computation, 1*, 221

Lieberoth, A., Pedersen, M. K., & Sherson, J. (2015). Play or science?: a study of learning and framing in crowdscience games. *Well Played, 4*, 30-55.

Magnussen, R., Hansen, S. D., Planke, T., & Sherson, J. F. (2014). Games as a platform for student participation in authentic scientific research, *EJEL, 12*, 259

Mekler, E. D., Brühlmann, F., Tuch, A. N., & Opwis, K. (2015). Towards understanding the effects of individual gamification elements on intrinsic motivation and performance. *Computers in Human Behavior*. doi:10.1016/j.chb.2015.08.048

Muntean, C. I. (2011). Raising engagement in e-learning through gamification. *The 6th International Conference on Virtual Learning ICVL*, 323-329.

Pedersen, M. K., Skyum, B., Heck, R., Müller, R., Bason, M. Lieberoth, A., & Sherson, J. F. (2016) Virtual Learning Enviroment for Interactive Engagement with Advanced Quantum Mechanics. *Phys. Rev. Phys. Educ. Res. 12*, 013102

Ryan, R. M. & Deci, E. L. (2000). Intrinsic and extrinsic motivations: Classic definitions and new directions. *Contemporary educational psychology, 25*, 54-67.

Vygotsky, L. S. (1978). *Mind in society: The development of higher psychological processes*. Cambridge, MA: Harvard University Press.

Wood, D., Bruner, J., & Ross, G. (1976). The role of tutoring in problem solving. *Journal of Child Psychology and Psychiatry, 17*, 89–100.